What is the effect of country-specific characteristics on the research performance of scientific institutions? Using multi-level statistical models to rank and map universities and research-focused institutions worldwide

Lutz Bornmann[1], Moritz Stefaner[2], Felix de Moya Anegón[3], & Rüdiger Mutz[4]

1       Division for Science and Innovation Studies, Administrative Headquarters of the Max Planck Society, Munich, Germany

2       Eickedorfer Damm 35, 28865 Lilienthal, Germany

3       CSIC, Institute of Public Goods and Policies (IPP), Madrid, Spain

4       Professorship for Social Psychology and Research on Higher Education, ETH Zurich, Zurich, Switzerland

**Abstract**

Bornmann, Stefaner, de Moya Anegón, and Mutz (2014) have introduced a web application (www.excellencemapping.net) which is linked to both academic ranking lists published hitherto (e.g. the Academic Ranking of World Universities) as well as spatial visualization approaches. The web application visualizes institutional performance within specific subject areas as ranking lists and on custom tile-based maps. The new, substantially enhanced version of the web application and the generalized linear mixed model for binomial data on which it is based are described in this paper. Scopus data were used which have been collected for the SCImago Institutions Ranking. Only those universities and research-focused institutions are considered that have published at least 500 articles, reviews and conference papers in the period 2006 to 2010 in a certain Scopus subject area. In the enhanced version, the effect of single covariates (such as the per capita GDP of a country in which an institution is located) on two performance metrics (best paper rate and best journal rate) is examined and visualized. A covariate-adjusted ranking and mapping of the institutions is produced in which the single covariates are held constant. The results on the performance of institutions can then be interpreted as if the institutions all had the same value (reference point) for the covariate in question. For example, those institutions can be identified worldwide showing a very good performance despite a bad financial situation in the corresponding country.

# 1      Introduction

International rankings have been published for around 10 years: in 2003, the Shanghai Jiao Tong University published the Academic Ranking of World Universities (ARWU) with The Times Higher Education QS Top University Ranking (THE QS) following a year later in 2004. "Despite significant volumes of criticism and commentary, and some boycotts by HEIs [Higher Education Institutions], rankings have become an increasingly popular way to compare higher education performance and productivity" (Hazelkorn, 2013, p. 9). In a list of the most prominent rankings Hazelkorn (2013) names 11 different international rankings (see also Rauhvargers, 2011). The most important source of data used for the various rankings are abstract and citation databases of peer-reviewed literature (primarily Scopus, which is provided by Elsevier and the Web of Science, WOS, from Thomson Reuters). Publication and citation data is used to make a statement about the productivity and the citation impact of institutions (Bornmann, de Moya Anegón, & Leydesdorff, 2012; Waltman et al., 2012).

Recent years have seen a number of different approaches which not only put institutions in a ranking list, but also show their performance on a map (Zhang, Perra, Gonçalves, Ciulla, & Vespignani, 2013). The advantage of this visualization is that regions and countries can be explored and searched for excellent institutions. However there is more than one benefit to using geography to show the institutions: Abramo, Cicero, and D'Angelo (2013) and Pan, Kaski, and Fortunato (2012) have found that geography has a relevant effect on the dynamics of science. This effect can be shown on a map (Bornmann & Waltman, 2011). There is an overview of the various approaches to mapping institutions in van Noorden (2010) and Frenken, Hardeman, and Hoekman (2009). In a recent study, Bornmann, Leydesdorff, Walch-Solimena, and Ettl (2011) presented methods to map centres of scientific excellence around the world. By colorizing cities worldwide according to the output of highly cited papers, their maps provide visualizations where cities with a high (or low) output of these papers can be found. Bornmann and Waltman (2011) follow their approach in general, but change the focus from mapping single cities to a more "sliding" visualization of broader regions. In a current study, Zhang, et al. (2013) "analyze the entire publication database of the American Physical Society generating longitudinal (50 years) citation networks geo-localized at the level of single urban areas."

Bornmann, Stefaner, et al. (2014) have introduced a web application (www.excellencemapping.net) which is linked to both spatial visualization approaches as well as academic ranking lists published hitherto. The web application visualizes institutional performance within specific subject areas as ranking lists and on custom tile-based maps. The web application is based on the results of a multilevel logistic regression. Multilevel models provide a very easy way to compare institutions, that is, whether they differ statistically significantly in their performance. The new, substantially enhanced version of the web application and the statistical analysis on which it is based are described in this paper. Multilevel models are used to test whether there are any systematic performance differences between institutions which justify their ranking.

In the enhanced version, the effect of single covariates (such as the per capita GDP of a country in which an institution is located) is examined and visualized. A covariate-adjusted ranking and mapping of the institutions is produced in which the single covariates are held constant. A number of studies have already shown that certain covariates can exert a significant influence on the results of rankings. In a secondary analysis of the Leiden Ranking (http://www.leidenranking.com) Bornmann, Mutz, and Daniel (2013) determined that the ranking of the universities changes significantly if context variables (such as the per capita GDP or the population of the country in which the university is located) are taken into account when the performance is measured. A study of German economic research institutes also finds that the ranking of these institutes changes with the inclusion of covariates (Ketzler & Zimmermann, 2013).

## 2 Methods

### 2.1 Data set

This study is based on Scopus data collected for the SCImago Institutions Ranking (http://www.scimagoir.com/). To obtain reliable data in terms of geo-coordinates (Bornmann, Leydesdorff, et al., 2011) and performance metrics (Waltman, et al., 2012), we only consider those institutions (universities or research-focused institutions) that have published at least 500 articles, reviews and conference papers in the period 2006 to 2010 in a certain Scopus subject area in the study.[1] Institutions with fewer than 500 papers in a category are not considered. Furthermore, only subject areas offered at least 50 institutions are included in the web application (Arts and Humanities, for example, is not included). We use this threshold in order to have sufficient institutions for a worldwide comparison. The full counting method was used (Vinkler, 2010) to attribute papers from the Scopus database to institutions: if an institution appears in the affiliation field of a paper, it is attributed to this institution (with a weight of 1). According to the results obtained by Waltman, et al. (2012), the overall correlation between a university ranking based on the full counting and fractional counting method is very high (r = .97). Furthermore, regression model specifications (see section 2.3) become very difficult in case of weighted data. The fractional counting method gives less weight (<1) to collaborative than to non-collaborative papers (= 1).

The citation impact of the publications from the institutions relates to the period from publication to mid-2013.

---

[1] The first version of the excellence mapping tool, Bornmann, Stefaner, et al. (2014) looked at the time period from 2005 to 2009. The first version of the excellence mapping tool can also be accessed in the current release. In the meantime, we have published the third version of the tool which refers to the period from 2007 to 2011. The third version is based on the same variables, methods, and techniques as the second version, which is described in this paper.

## 2.2 Variables

*Dependent variables*

The performance of the institutions is measured with two indicators in this study. In line with the recommendation by Rodríguez-Navarro (2012) for university research assessments, both indicators reflect "substantive contributions made to scientific progress." The first indicator, called the <u>best paper rate</u> (Bornmann, et al., 2012), shows the proportion of publications from an institution which belong to the 10% most cited publications in their subject area and publication year. According to Waltman, et al. (2012) the best paper rate (named as $PP_{top\ 10\%}$) is "the most important impact indicator" (p. 2425) for the ranking of universities by research performance. We used this indicator in the first version of the excellence mapping tool (Bornmann, Stefaner, et al., 2014).

To identify the publications of an institution belonging to the 10% most cited publications in their subject area and publication year, the citations $X_i$ that were received by the ith papers within n papers published in a given subject area (and publication year) were gathered. Then the papers were ranked in ascending order

$X_1 \leq X_2 \leq \ldots \leq X_n,$

where $X_1$ and $X_n$ denote the number of citations received respectively by the least and most cited paper. Where citation counts are equal, the SCImago Journal Rank SJR2 indicator (Guerrero-Bote & de Moya-Anegon, 2012) of the journal which published the papers is used as a second sort key (from highest to lowest). This journal metric takes into account not only the prestige of the citing scientific publication but also its closeness to the cited journal. Finally, in each subject area and publication year, each individual publication was assigned a percentile rank based on this distribution. If, for example, a single paper within a subject area had 50 citations, and this citation count was equal to or greater than the citation counts of 90% of all papers in the subject area, then the percentile rank of this paper would be 90. The paper would be in the 90$^{th}$ percentile and would belong to the 10% most cited papers within the subject area. There are different approaches available for calculating percentile-based indicators (see an overview in Bornmann, Leydesdorff, & Mutz, 2013; Waltman & Schreiber, 2013). In a comparison, the approach used for the SCImago Institutions Ranking proved to have particular advantages over others (Bornmann, Leydesdorff, & Wang, 2013; Bornmann, Leydesdorff, & Wang, 2014).

The best paper rate of the institutions included in this study is on average 15% (see Table 1). The proportion of the institutions' papers belonging to highly-cited papers is therefore on average 5 percentage points higher than the expected proportion (10%). Across the institutions, the proportion ranges from 0% (minimum) to 56% (maximum).

A second excellence indicator (not integrated in the first version) is taken into account in the new release of the excellence mapping tool: the ratio of papers that an institution publishes in the most influential scholarly journals of the world (<u>best journal rate</u>). The most influential journals are those which ranked in the first quartile (25%) of their subject areas (journal sets) as ordered by the SCImago Journal Rank SJR indicator (Gonzalez-Pereira, Guerrero-Bote, & Moya-Anegon, 2010). While the best paper rate gives information about the long-term success of an institution's publications, the best journal rate describes an earlier stage in the process, the ability of an institution to publish its research results in reputable journals. As Table 1 shows, on average the institutions have published around half of their papers in the most influential scholarly journals of the world. A $\text{Mean}_{\text{best journal rate}}=.56$ indicates that on average, the institutions are performing better than the expected value of 25%. This finding might partly result from the full counting method used in this study. Highly cited publications and publications in high-impact journals tend to be the ones that have a relatively large number of authors, typically from different institutions. Because in the full counting approach these publications are counted multiple times, once for each of the co-authoring institutions, they create the situation in which most institutions are performing above expectation. This would be quite different in the case of the fractional counting approach, where it is mathematically impossible for everyone to perform above expectation (i.e. that means above what would be expected, if each publication was counted only once).

Table 2 shows the number of institutions which are considered as datasets for the 17 subject areas in this study. Out of the 27 available subject areas in Scopus, only those are selected for the study which include at least 50 institutions worldwide. For example, 511 institutions within the subject area of chemistry were included in the analyses. The mean best paper rate for these institutions is .13 (13%) and the mean best journal rate .62 (62%).

*Covariates*

To be able to explain the performance differences among the institutions in the regression model, we included the following variables as covariates (see Table 1). The covariates are also utilized to create a covariate-adjusted ranking of the institutions:

(1) Proportion of papers from one institution which were produced in an international collaboration (international collaboration).

(2) Corruption perception index.

(3) Number of residents in a country (number of residents).

(4) Gross domestic product (GDP) per capita of a country (gross domestic product).

While the international collaboration covariate relates to individual institutions, all the other covariates apply at the level of individual countries.

(1): As Table 1 shows, averaged across all the institutions included in this study, 35% of the publications (of an institution) were produced in an international collaboration. The percentage of papers in international collaboration is computed by analyzing an institution's output whose affiliations include more than one country address. While the minimum value of an institution for international collaboration is 0%, the maximum is 98%. As the overview from Lancho-Barrantes, Guerrero-Bote, and Moya-Anegón (2013) on studies which have investigated the influence of collaborations on the citation impact of publications shows, we can expect publications produced in an international collaboration to have more impact than those which were not. The current study by Adams (2013) also arrives at this result.

(2): The corruption perception index which is measured for 85 countries is based on surveys of primarily business people and risk analysts who might adequately recognize corruption when they see it (see http://cpi.transparency.org/cpi2012/results/). The score ranges from 0 ("country's business transactions are almost all dominated by bribery") to 100 ("highly clean") (Wilhelm, 2002). The mean corruption perception index averaged over all the countries in which at least one of the institutions included in this study is located is 54.1 (roughly equivalent to that of Hungary). Venezuela has the lowest value at min = 19 and Denmark the highest at max = 90. Although we are not aware of any studies investigating the influence of corruption in a country on the performance of its research institutions, we assume that it would be negative. For example, the results of Ariu and Squicciarini (2013) show that "in countries where corrupt people – through family ties, money or political affiliations – determine access to the job market, emigration of highly skilled labor is high and immigration of foreign talents is reduced, thus creating a net deficit" (p. 504).

(3): The number of residents in each country was researched in the report issued by the Deutsche Stiftung Weltbevölkerung (http://www.weltbevoelkerung.de). The number of residents also varies greatly across countries: The number ranges from 0.3 to 1,350.4 (million). Of course, the countries' output is determined by their population size (Harzing & Giroud, 2014). As a larger population usually also means a larger pool of potential (excellent) scientists, we therefore expect positive correlations with citation impact. The results of Bornmann, Mutz, et al. (2013) show that this covariate does indeed have a statistically significant effect on the universities' performance in the Leiden Ranking: The larger the number of inhabitants, the higher the citation impact of a university. According to results produced by Mazloumian, Helbing, Lozano, Light, and Börner (2013) the citation impact of papers from a country is related to the size of a country.

(4): The gross domestic product (GDP) per capita is the gross domestic product converted to international dollars (www.worldbank.org). Regarding GDP we assume – and this has also been confirmed by Bornmann, Mutz, et al. (2013) with the aid of data from the Leiden Ranking – that in a country where more money is available generally, there will be more funding for research, and thus higher-level research can be conducted than in countries with limited financial means. More results concerning available money in a country and research performance can be found in Meo, Al Masri, Usmani, Memon, and Zaidi (2013), Allik (2013), and Grossetti et al. (2013). Miranda and Lima (2010) point out that "the knowledge evolution, as seen through the evolution of major scientific discoveries and impacting technological inventions, is exponentially correlated to the GDP" (p. 92). As Table 1 shows, GDP in the countries included in this study ranges from 479 to 99,143; the median value is 14,501. GDP for 2011 was researched at the World Bank (http://data.worldbank.org/) and in the CIA world fact book (www.cia.gov/library/publications/the-world-factbook/).

### 2.3 Statistical model

As in our paper introducing the first version of the excellence mapping tool, we prefer here a generalized linear mixed model (GLM) for binomial data, which takes into account the hierarchical structure of data and properly estimates the standard errors (Bornmann, Stefaner, et al., 2014; Mutz & Daniel, 2007). The model can handle complex data structure with a small set of parameters. For instance, one parameter is sufficient to test statistically, whether the institutional performances vary only by random chance (i.e. as random samples of the same population) or in a systematic manner. Rankings among institutions only make sense if there are systematic differences between these institutions.

Our data are organized as follows: For each institution (which is the group variable), the total number of papers, denoted $n_i$, the number of papers which belong to the 10% most cited publications (best paper rate), the number of papers that are published in the most influential scholarly journals (best journal rate), and institution-specific covariates (e.g. GDP) are available. If $Y_i$ denotes the institutional number of papers, i, which belong to the 10% most cited papers/ are published in the most influential journals, this is an independent binomial random variable (Schabenberger, 2005). To model the best paper and best journal rates probabilities, $p_i$, for each subject area, one needs to account for the fixed effect of the covariate and the random effects of the institutions. The following model is assumed to relate the fixed and random effects model to the logit of the probability $p_i$ (Bornmann, Mutz, Marx, Schier, & Daniel, 2011; Bornmann, Stefaner, et al., 2014; Hox, 2010; Schabenberger, 2005):

$$\log\left\{\frac{p_i}{1-p_i}\right\} = \beta_0 + \beta_1 x_i + u_i \qquad (1)$$

We assume that the parameters $\beta_0$, $\beta_1$, and $u_i$ are not correlated with the covariate $x_i$, $cov(x_i, u_i) = 0$. Whereas $\beta_0$ and $\beta_1$ are fixed effects, $u_i$ are the random effects. We understand the 500 institutions (with a substantial publication output), which we selected for this study at a certain time point, as a sample of all institutions (with a substantial output) across a certain time interval (between two releases of institutional data). As the assumption of random sampling is questionable for our data, we adopted the concept of exchangeability from Raudenbush (1993, p. 330). In a Bayesian perspective, exchangeability reflects the investigator's uncertain state of knowledge about a large number of parameters with favourable consequence of fostering the efficiency of the estimation by using multilevel models.

Because the covariates used in this study are scaled very differently (e.g. people, $), they were z-transformed with a mean value of 0 and a standard deviation of 1.0 (Hox, 2010). The transformation does not change the model fit and has no influence on the institutional

rankings. The standardization has the following advantages over non-standardized covariates: (1) The intercept in the statistical model, where the regression line crosses the y-axis, becomes a meaningful value: It represents the value of a fictitious institution for which a covariate is exactly average. (2) The different models can be simpler compared. (3) Due to different scales of the covariates the regression parameters become very small, if covariates are not standardized. (4) Standardization helps to avoid convergence problems and to speed up the numerical calculation of the models.

The conditional distribution of $Y_i$ (binomial) and the distribution of the random effects (normal) are specified as follows (Schabenberger, 2005, p. 4):

$$Y_i|u_i \sim \text{binomial}(n_i, p_i) \qquad (2)$$
$$u_i \sim \text{iid } N(0,\sigma^2_u),$$

where $\sigma^2_u$ is the variance of the random effects $u_i$.

There is a so-called intra-class correlation conditioned on the covariate x between papers within institutions with $\rho = \sigma^2_{u0|x}/(3.29+\sigma^2_{u0|x})$ which reflects the homogeneity of papers within institutions given the covariate x. The Wald test can test whether $\sigma^2_{u0|x}$ deviates from 0 (the null hypothesis). If the Wald test is statistically significant, there are systematic differences between institutions beyond random fluctuations with respect to the best paper and best journal rate, even if these differences are controlled for a covariate. Comparisons or rankings of institutions only make sense in this case. If on the one hand a covariate does explain less of the random intercept variance across institutions, the ranking of institutions remains the same with or without it. On the other hand, if a covariate explains almost all of the random intercept variance, comparisons among institutions are no longer reasonable. Then, the level-2 residuals reflect only random differences among the institutions. Due to the fact that only cluster-covariates and not paper-based covariates are used, the model parameters need not be adjusted and can be compared across models (Bauer, 2009).

The proportion of variance $R^2_{u.x}$ explained by the cluster-covariates is the difference between the random intercept variance $\sigma^2_{u0}$, if no covariate is included, and $\sigma^2_{u0|x}$, if the covariate x is included in the model, divided by $\sigma^2_{u0}$: $R^2_{u.x} = (\sigma^2_{u0} - \sigma^2_{u0|x})/\sigma^2_{u0}$.

Most importantly, the multilevel model can be used to estimate so-called Empirical Bayes (EB) or shrinkage estimates which are more precise than their empirical counterparts, the raw probabilities (Bornmann, Stefaner, et al., 2014; Hox, 2010; SAS Institute Inc., 2008). The EB and the confidence intervals can be transformed back to probabilities to facilitate the interpretation of the results. The multiplication of standard errors by 1.39 instead of 1.96 results in so-called Goldstein-adjusted confidence intervals (Goldstein & Healy, 1994) with the property that if the confidence intervals of two institutions do not overlap, they differ statistically significantly ($\alpha = 5\%$) in their estimates (i.e. best paper and best journal rates probabilities). If the 95% confidence interval does not include the mean best paper or best journal rate across all institutions, the authors located at this institution have published a statistically significantly greater or smaller number of highly cited papers or of papers in reputable journals than the average across all institutions. In the case of a Goldstein-adjusted confidence interval, this test can only be done on the 16.3% probability level, rather than on the usual 5% level (Bornmann, Stefaner, et al., 2014).

To generate the data for the visualization for each of the 17 subject areas a multilevel analysis was calculated for the best paper rate as well as for the best journal rate. The intercept of the model without covariates provides for the reference value ($\beta_0$) to calculate the residuals $r_j$ in terms of probabilities ranging from 0 to 1 comparable to the original data: $r_j = \exp(\beta_0 + u_{0j})/(1 + (\beta_0 + u_{0j}))$. The choice of the reference point ($\beta_0$) is arbitrary and does not change the ranking of the institutions.

It would go far beyond this paper to report the results of all 2 * 17 analyses. Therefore, we present two overall models (see Table 3 and Table 4), in which the subject area and the interaction subject area × covariate are additionally included as fixed effects in the multilevel model (Eq. 1) using dummy coding. Due to the fact that there are 16 non-redundant pieces of information for 17 subject areas, 16 main effects (subject area) and 16 interactions (subject area × covariate) with corresponding parameters ($\beta_2$ - $\beta_{18}$, $\beta_{19}$ - $\beta_{35}$) were added to the fixed-effects model (Eq. 1) as follows:

$$\log\left\{\frac{p_i}{1-p_i}\right\} = \beta_0 + \beta_1 x_i + \beta_2 d_{1i} + ... + \beta_{18} d_{16i} + \beta_{19} d_{1i} x_i + ... + \beta_{35} d_{16i} x_i + u_i \quad (3)$$

Instead of reporting all the estimates of the 32 parameters, only the overall test results (F- tests) were reported in Table 3 and Table 4.

The more appropriate way to analyze the data would be to include the subject area as an additional third level in the multilevel analysis. Unfortunately, such complex models with random intercepts and random slopes for subject areas could not be estimated. Also, the low number of subject areas (n=17) does not justify the inclusion of a third level.

The analyses were done using the proc glimmix procedure implemented in the SAS statistical software (SAS Institute Inc., 2008; Schabenberger, 2005).

**2.4    Interactive Visualization**

We implemented an interactive statistical mapping application, which presents the bibliometric data in the form of a world map and table view. Both forms allow the data to be examined by subject area, covariate, and data set edition (publications years 2005 to 2009 and 2006 to 2010). The user can choose various institutions for direct comparison. The application was implemented using modern web technologies and Open Street Map[2] data provided through MapBox[3]. It is based on the javascript frameworks backbone.js[4], jquery[5] and d3.js[6].

## 3 Results

### 3.1 Results of the multilevel regression analysis

*Overall systematic institutional differences*

A model was taken as the starting point in the statistical analysis, which only includes a random and fixed intercept and subject area as fixed factors (Table 3, Table 4, model $M_0$). The random intercept variance $\sigma^2_{u0|x}$ represents the variability of random intercepts across institutions whereas the data is controlled for the main effect of the subject area (differences between subject areas are discarded). The parameter estimates show that there is sufficient residual variability left ($\sigma^2_{u0|x} = 0.50$ for best paper rate and $\sigma^2_{u0|x} = 0.78$ for best journal rate) to justify an institutional ranking. Given the null hypothesis of zero variance in the population, the probability of the empirical results for the variance component $\sigma^2_{u0|x}$ is below 0.05, i.e. the result is statistically significant. The intra-class correlation [0; 1] as a measure of the variability between institutions to the total variability as sum of the variability between and within institutions (random fluctuations) amounts to 0.13 for best paper rate and 0.18 for best journal rate. A statistically significant F-test for the subject area shows that there are statistically significant differences between the subject areas. It is therefore reasonable to separate rankings for subject areas, which was implemented in our web application.

*Impact of covariates*

For each covariate a multilevel model was estimated for the best paper rate as well as for the best journal rate (Table 3, Table 4). Beside the correlation between the corruption perception index and the GDP (r =.81) the correlations among the covariates are small or medium (Table 5). Therefore, a separate analysis for each covariate can be justified. In all models with covariates there is sufficient residual variance $\sigma^2_{u0|x}$ left in order to justify a covariate-specific ranking. The null hypothesis of zero variance in the population can always be rejected with p<.05. Due to the fact that the same model (random-intercept fixed-effects model) with different covariates is used, comparison of models using information criteria (Schwarz Bayesian criterion, BIC) or deviance is actually not appropriate. However, they can give some hints for optimal models.

---

[2] http://www.openstreetmap.org
[3] https://www.mapbox.com
[4] http://backbonejs.org
[5] http://jquery.org
[6] http://d3js.org

    The results of the models for the best paper rate are comparable to the results of the models for the best journal rate, so we will discuss the results for the best paper rate only in the following. In terms of the amount of explained variance, GDP is the most important factor, which explains with $R^2_{u.cov} =.47$ about 47% of the random-intercept variance for the best paper rate, followed by the corruption perception index ($R^2_{u.cov} =.41$, 41%) and the international collaboration ($R^2_{u.cov} =.32$, 32%). Except for the number of residents in a country the effect of the covariates on the best paper rate is positive: The higher the value of the covariate, the higher the best paper rate. But the higher the number of residents, the lower the best paper rate. The statistically significant F-tests for the interaction subject area × covariate indicate that the impact of all covariates is different across the subject areas, which justifies a separate multilevel analysis for each subject area. Figure 1 shows the predicted best paper rates in dependence on GDP for different subject areas (without considering differences between institutions). The best paper rate increases exponentially with increasing GDP with large differences among the 17 subject areas. These differences are particularly visible where the GDP exceeds the threshold of $ 60,000. Whereas the best paper rate in medicine strongly depends on the GDP, the best paper rate in mathematics is less strongly associated.

### 3.2 The Web Application

The URL of the web application is: http://www.excellencemapping.net. There is a very short description of the visualization displayed in the upper right section of the web application with a link to click for "More information". The page with the detailed description includes the affiliations of the authors of the web application and a link to this research paper. Under the short description of the web application, the user can choose whether to view the first release of our web application, which relates to the publication period 2005-2009, or the second with the publication period 2006-2010. Only in the second release, the user can select two different excellence indicators and several covariates. In the upper right section of the web application, the user can select from 17 subject areas for the visualization. Under the selection window for the subject area, there is another for the covariate (for selecting the corruption perception index, for example).

If the user selects a covariate, the probabilities of (i) publishing in the most influential journals (best journal rate) or (ii) publishing highly cited papers (best paper rate) is displayed adjusted (controlled) for the selected covariate. The results on the performance of institutions can then be interpreted as if the institutions all had the same value (reference point) for the covariate in question. Each covariate was z-transformed over the whole data set (with M=0 and S=1), so that the average probability shows the value in which the covariate in question has the value 0, i.e. exactly equivalent to the median. This allows the results of the model with and without the covariates to be compared.

Below the selection windows for the subject area and the covariates, users can select one of the two excellence indicators (best paper rate or best journal rate). Our tool shows for each of these indicators the residues from the regression model (random effects) converted to probabilities. In order to have values on the original scale for both indicators for the tool (i.e. proportion of papers in the excellent range or published in the best journals), the intercept was added to the residues. Users can tick "Show statistically significant results only" to reduce the set of visualized institutions in a field to only those which differ statistically significantly in their performance from the mean value.

The map on the left-hand side of the screen shows a circle for each institution with a paper output greater than or equal to 500 for a selected subject area (e.g. Physics and Astronomy). Users can move the map to different regions with the mouse (click and drag) and zoom in (or out) with the mouse wheel. Country and city labels and map details appear only at zoom levels of a certain depth, primarily in order to facilitate perception of the data markers. Zooming can also be done with the control buttons at the top left of the screen. The circle area for each institution on the map is proportional to the number of published papers in the respective subject area. For example, the Centre National de la Recherche Scientifique (CNRS) has the largest circle (in Europe) on the Physics and Astronomy map, highlighting the high output of papers in this subject area. As several circles overlap on larger cities, users can select all the circles in a certain region with the mouse, by holding down the shift key and marking out the area on the map in which the institutions in question are located. These institutions are then displayed on the right-hand side of the web application under "Your selection". The color of the circles on the map indicates the excellence indicator value for the respective institution using a diverging color scale, from blue through grey to red (without any reference to statistical testing): If the excellence indicator value for an institution is greater than the mean (expected) value across all institutions, its circle has a blue tint. Circles with red colors mark institutions with excellence indicator values lower than the mean. Grey circles indicate a value close to the expected value.

All those institutions which are taken into account in the multi-level model for a subject area (section "Institutional scores") are listed on the right-hand side of the web application. The name, the country, and the number of all the papers published ("Papers") are displayed for each institution. In addition, the probabilities of (i) publishing in the most influential journals (best journal rate) or (ii) publishing highly cited papers (best paper rate) are visualized ("Indicator value"). The greater the confidence interval of the probability, the more unreliable for an institution it is. If the confidence interval does not overlap with the mean proportion across all institutions (the mean is visualized by the short line in the middle of "Indicator value"), this institution has published a statistically significantly higher (or lower) best paper or best journal rate than the average across all the institutions ($\alpha = 0.165$). The institutions in the list can be sorted (in descending or ascending order in the case of numbers) by clicking on the relevant heading. Thus, the top or worst performers in a field can be identified by clicking on "Indicator value." Clicking on "Papers" puts the institutions with high productivity in terms of paper numbers at the top of the list (or at the end). In Biochemistry, Genetics and Molecular Biology, for example, the institution with the highest productivity between 2006 and 2010 is the CNRS; in terms of the best paper rate, the best-performing institution is the Broad Institute of MIT and Harvard. The column farthest on the right ("Δ rank") in the "Institutional Scores" section shows for each institution by how many rank places it goes up (green, arrow pointing upwards) or goes down (red, arrow pointing downwards), if the user selects a certain covariate. For example, the Institute for High Energy Physics (RUS) improves its position by 16 places compared to the ranking which does not take the covariate "corruption perception index" into account in the Physics and Astronomy subject area. The ranking differences in this column always relate to all the institutions included. The differences do not therefore change if one looks at only the statistically significant results.

If a covariate has been chosen, one can, for example, sort the institutions by Δ rank, which puts the institutions which benefit most from the covariate being taken into account at the top of the list. Using the search field at the top right, the user can find a specific institution in the list. To identify the institutions for a specific country, click on "Country". Then the institutions are first sorted by country and second by the indicator value (in ascending or descending order). "Your selection" is intended to be the section for the user to compare institutions of interest directly. If the confidence intervals of two institutions under "Indicator value" do not overlap, they differ statistically significantly on the 5% level in the best paper or best journal rate. For example, in Physics and Astronomy, Stanford University and the Helmholtz Gemeinschaft are visualized without overlap (publication years 2006 to 2010). The selected institutions in "Your selection" can be sorted by each heading in different orders. These institutions are also marked on the map with a black border. Thus, both institutional lists and institutional maps are linked by the section "Your selection". For the comparison of different institutions, it is not only possible to select them in the list but also on the map with a mouse click. A new comparison of institutions can be started by clicking on "Clear".

If the user has selected some institutions or has sorted them in a certain order, the selection and sort order are retained if the subject area is changed. This feature makes it possible to compare the results for certain institutions across different subject areas directly.

## 4      Discussion

According to Hazelkorn (2013) we can break the development in university rankings down into three phases: In Phase 1 primarily private universities in the USA were compared with each other with the aid of indicators such as faculty expertise, graduate success in later life, and academic resources. Phase 2 began around the end of the 1960s, initially in the USA and later in other countries, when more and more national rankings with every institution were set up. These rankings were already often based on bibliometric data from the newly developed Science and Social Sciences Citation Indexes (Garfield, 1955). The era of global rankings, the third phase, came at the beginning of the 21st century with the publication of the ARWU (see above). Since then a host of other global rankings have been developed in addition to ARWU. We can view the U-Multirank commissioned by the European Union as the current end point of this development (van Vught & Ziegele, 2012). This comprehensive ranking "compares the performance of universities and colleges not only in research, but also in teaching, knowledge transfer, international orientation and regional engagement; it is a multi-dimensional ranking. It presents performance profiles for universities across the five dimensions using a broad range of performance indicators" (CHE Centre for Higher Education). The compilation of the data for the ranking means a high workload for the involved institutions.

With the excellence mapping introduced in this paper, we are taking a different approach to designing a ranking than that chosen for U-Multirank, for example. In our web application, we have limited ourselves to the most important dimension in measuring research performance – the identification of research excellence – and created a study which has enabled institutions to be not only ranked, but also mapped by using the most suitable indicators and modern statistical processes. While the ranking allows a direct comparison of the institutions regarding the quality of the journal in which their papers appear (best journal rate) and of the publication output (best paper rate, measured using the citation impact), mapping the institutions allows the research performance of institutions in larger regions throughout the world to be explored. Compared to the mapping and ranking approaches introduced hitherto, our underlying statistics (multi-level models) are analytically oriented – following Bornmann and Leydesdorff (2011) – by allowing (1) the estimation of statistically more appropriate values for the best paper and best journal rates than the observed values; (2) the calculation of confidence intervals as reliability measures for the institutional performance; (3) the comparison of a single institution with an "average" institution in a subject area, and (4) the direct comparison of at least two institutions. (5) Furthermore, taking covariates into account when mapping and ranking institutions allows an adjusted view of institutional research performance. The covariates make a what-if-analysis possible. For example, with our application it is possible to look at the performance of institutions worldwide as if they were located in a country with the same financial background (that is, the same GDP).

Informed peer review is seen as the ideal way forward for the evaluation of research units (such as scientists, research groups or institutions) (Bornmann, 2011). Experts working in the same discipline as the unit being evaluated make an assessment of its research performance. This evaluation can only be undertaken by experts from the same subject area as only they and no others have the necessary knowledge. However, peer review comes up against its limits when a large number of research units need to be evaluated which are also geographically distributed over a wide area. "For obvious reasons of costs and time it is clearly unthinkable to utilize peer-review to evaluate the entire output of a national research system" (Abramo, et al., 2013). One must rely on quantitative indicators for an overview of institutions throughout the world – particularly indicators based on bibliometric data. Unlike other data (such as data on third party funding), bibliometric data is available all over the world and in a comparable form for every institution. It can also be evaluated for a period of time and across subject areas (with the appropriate normalization).

The excellence mapping tool presented in this paper is the second version of a web application for the visualization of institutional research performance. We plan to continue to develop the application over the next few years. The development will depend on the data to be used: more up-to-date data will be included, but the older data will be retained. Furthermore, new indicators and statistical analyses will be included.

**Acknowledgements**

We would like to thank two anonymous reviewers for their valuable feedback to improve the paper.

**References**

Abramo, G., Cicero, T., & D'Angelo, C. (2013). National peer-review research assessment exercises for the hard sciences can be a complete waste of money: the Italian case. *Scientometrics, 95*(1), 311-324. doi: 10.1007/s11192-012-0875-6.

Adams, J. (2013). Collaborations: the fourth age of research. *Nature, 497*(7451), 557-560. doi: 10.1038/497557a.

Allik, J. (2013). Factors affecting bibliometric indicators of scientific quality. *Trames - Journal of the Humanities and Social Sciences, 17*(3), 199-214.

Ariu, A., & Squicciarini, M. P. (2013). The balance of brains - corruption and migration. *EMBO Reports, 14*(6), 502-504. doi: 10.1038/embor.2013.59.

Bauer, D. J. (2009). A note on comparing the estimates of models for cluster-correlated or longitudinal data with binary or ordinal outcomes. *Psychometrika, 74*(1), 97-105. doi: 10.1007/S11336-008-9080-1.

Bornmann, L. (2011). Scientific peer review. *Annual Review of Information Science and Technology, 45*, 199-245.

Bornmann, L., de Moya Anegón, F., & Leydesdorff, L. (2012). The new Excellence Indicator in the World Report of the SCImago Institutions Rankings 2011. *Journal of Informetrics, 6*(2), 333-335. doi: 10.1016/j.joi.2011.11.006.

Bornmann, L., & Leydesdorff, L. (2011). Which cities produce more excellent papers than can be expected? A new mapping approach—using Google Maps—based on statistical significance testing. *Journal of the American Society of Information Science and Technology, 62*(10), 1954-1962.

Bornmann, L., Leydesdorff, L., & Mutz, R. (2013). The use of percentiles and percentile rank classes in the analysis of bibliometric data: opportunities and limits. *Journal of Informetrics, 7*(1), 158-165.

Bornmann, L., Leydesdorff, L., Walch-Solimena, C., & Ettl, C. (2011). Mapping excellence in the geography of science: an approach based on Scopus data. *Journal of Informetrics, 5*(4), 537-546.

Bornmann, L., Leydesdorff, L., & Wang, J. (2013). Which percentile-based approach should be preferred for calculating normalized citation impact values? An empirical comparison of five approaches including a newly developed citation-rank approach (P100). *Journal of Informetrics, 7*(4), 933-944. doi: 10.1016/j.joi.2013.09.003.

Bornmann, L., Leydesdorff, L., & Wang, J. (2014). How to improve the prediction based on citation impact percentiles for years shortly after the publication date? *Journal of Informetrics, 8*(1), 175-180.

Bornmann, L., Mutz, R., & Daniel, H.-D. (2013). A multilevel-statistical reformulation of citation-based university rankings: the Leiden Ranking 2011/2012. *Journal of the American Society for Information Science and Technology, 64*(8), 1649-1658.

Bornmann, L., Mutz, R., Marx, W., Schier, H., & Daniel, H.-D. (2011). A multilevel modelling approach to investigating the predictive validity of editorial decisions: do the editors of a high-profile journal select manuscripts that are highly cited after publication? *Journal of the Royal Statistical Society - Series A (Statistics in Society), 174*(4), 857-879. doi: 10.1111/j.1467-985X.2011.00689.x.

Bornmann, L., Stefaner, M., de Moya Anegón, F., & Mutz, R. (2014). Ranking and mapping of universities and research-focused institutions worldwide based on highly-cited papers: A visualization of results from multi-level models. *Online Information Review, 38*(1), 43-58.

Bornmann, L., & Waltman, L. (2011). The detection of "hot regions" in the geography of science: a visualization approach by using density maps. *Journal of Informetrics, 5*(4), 547-553.

CHE Centre for Higher Education. *U-Multirank - key questions and answers*. Gütersloh, Germany: CHE Centre for Higher Education.

Frenken, K., Hardeman, S., & Hoekman, J. (2009). Spatial scientometrics: towards a cumulative research program. *Journal of Informetrics, 3*(3), 222-232. doi: 10.1016/j.joi.2009.03.005.

Garfield, E. (1955). Citation indexes for science - new dimension in documentation through association of ideas. *Science, 122*(3159), 108-111.

Goldstein, H., & Healy, M. J. R. (1994). The graphical representation of a collection of means. *Journal of the Royal Statistical Society Series a-Statistics in Society, 158*, 175-177.

Gonzalez-Pereira, B., Guerrero-Bote, V. P., & Moya-Anegon, F. (2010). A new approach to the metric of journals' scientific prestige: the SJR indicator. *Journal of Informetrics, 4*(3), 379-391. doi: 10.1016/j.joi.2010.03.002.

Grossetti, M., Eckert, D., Gingras, Y., Jégou, L., Larivière, V., & Milard, B. (2013). Cities and the geographical deconcentration of scientific activity: A multilevel analysis of publications (1987–2007). *Urban Studies*. doi: 10.1177/0042098013506047.

Guerrero-Bote, V. P., & de Moya-Anegon, F. (2012). A further step forward in measuring journals' scientific prestige: the SJR2 indicator. *Journal of Informetrics, 6*(4), 674-688.

Harzing, A.-W., & Giroud, A. (2014). The competitive advantage of nations: an application to academia. *Journal of Informetrics, 8*(1), 29-42. doi: http://dx.doi.org/10.1016/j.joi.2013.10.007.

Hazelkorn, E. (2013). Reflections on a decade of global rankings: what we've learned and outstanding issues. *Beiträge zur Hochschulforschung, 35*(2), 8-33.

Hox, J. J. (2010). *Multilevel analysis: techniques and applications* (2nd ed.). New York, NY, USA: Routledge.

Ketzler, R., & Zimmermann, K. (2013). A citation-analysis of economic research institutes. *Scientometrics, 95*(3), 1095-1112. doi: 10.1007/s11192-012-0850-2.

Lancho-Barrantes, B., Guerrero-Bote, V., & Moya-Anegón, F. (2013). Citation increments between collaborating countries. *Scientometrics, 94*(3), 817-831. doi: 10.1007/s11192-012-0797-3.

Mazloumian, A., Helbing, D., Lozano, S., Light, R. P., & Börner, K. (2013). Global multi-level analysis of the 'Scientific Food Web'. *Scientific Reports, 3*. doi: http://www.nature.com/srep/2013/130130/srep01167/abs/srep01167.html#supplementary-information.

Meo, S. A., Al Masri, A. A., Usmani, A. M., Memon, A. N., & Zaidi, S. Z. (2013). Impact of GDP, spending on R&D, number of universities and scientific journals on research publications among Asian countries. *PLoS ONE, 8*(6), e66449. doi: 10.1371/journal.pone.0066449.

Miranda, L. C. M., & Lima, C. A. S. (2010). On trends and rhythms in scientific and technological knowledge evolution: a quantitative analysis. *International Journal of Technology Intelligence and Planning, 6*(1), 76-109109.

Mutz, R., & Daniel, H. D. (2007). Development of a ranking procedure by mixed Rasch model and multilevel analysis - psychology as an example. *Diagnostica, 53*(1), 3-16. doi: Doi 10.1026/0012-1924.53.1.3.

Pan, R. K., Kaski, K., & Fortunato, S. (2012). World citation and collaboration networks: uncovering the role of geography in science. *Sci. Rep., 2*. doi: http://www.nature.com/srep/2012/121129/srep00902/abs/srep00902.html#supplementary-information.

Raudenbush, S. W. (1993). A crossed random effects model for unbalanced data with applications in cross-sectional and longitudinal research. *Journal of Educational Statistics, 18*(4), 321-349. doi: 10.2307/1165158.

Rauhvargers, A. (2011). *Global university rankings and their impact*. Brussels, Belgium: European University Association (EUA).

Rodríguez-Navarro, A. (2012). Counting highly cited papers for university research assessment: conceptual and technical issues. *PLoS ONE, 7*(10), e47210. doi: 10.1371/journal.pone.0047210.

SAS Institute Inc. (2008). *SAS/STAT® 9.2 user's guide*. Cary, NC, USA: SAS Institute Inc.

Schabenberger, O. (2005). Introducing the GLIMMIX procedure for generalized linear mixed models. In SAS Institute Inc. (Ed.), *Proceedings of the thirtieth annual SAS® users group international conference* (pp. 196-130). Cary, NC, USA: SAS Institute Inc.

van Noorden, R. (2010). Cities: building the best cities for science. Which urban regions produce the best research - and can their success be replicated? *Nature, 467*, 906-908. doi: 10.1038/467906a.

van Vught, F. A., & Ziegele, F. (Eds.). (2012). *Multidimensional ranking: the design and development of U-Multirank*. Dordrecht, the Netherlands: Springer.

Vinkler, P. (2010). *The evaluation of research by scientometric indicators*. Oxford, UK: Chandos Publishing.

Waltman, L., Calero-Medina, C., Kosten, J., Noyons, E. C. M., Tijssen, R. J. W., van Eck, N. J., . . . Wouters, P. (2012). The Leiden Ranking 2011/2012: data collection, indicators, and interpretation. *Journal of the American Society for Information Science and Technology, 63*(12), 2419-2432.

Waltman, L., & Schreiber, M. (2013). On the calculation of percentile-based bibliometric indicators. *Journal of the American Society for Information Science and Technology, 64*(2), 372-379.

Wilhelm, P. G. (2002). International validation of the Corruption Perceptions Index: implications for business ethics and entrepreneurship education. *Journal of Business Ethics, 35*(3), 177-189. doi: Doi 10.1023/A:1013882225402.

Zhang, Q., Perra, N., Gonçalves, B., Ciulla, F., & Vespignani, A. (2013). Characterizing scientific production and consumption in Physics. *Sci. Rep., 3*. doi: 10.1038/srep01640.

Table 1. Sample description

| Variable | N | Mean | SD | MIN | 25% | Med | 75% | MAX |
| --- | --- | --- | --- | --- | --- | --- | --- | --- |
| *Institution Level* | | | | | | | | |
| Best paper rate | 6,988 | 0.15 | 0.07 | 0 | 0.10 | 0.15 | 0.20 | 0.56 |
| Best journal rate | 6,988 | 0.56 | 0.19 | 0 | 0.44 | 0.60 | 0.70 | 0.98 |
| International collaboration | 6,988 | 0.35 | 0.17 | 0 | 0.22 | 0.34 | 0.47 | 0.98 |
| *Country Level* | | | | | | | | |
| Corruption perception index | 73 | 54.1 | 20.4 | 19 | 37 | 49 | 72 | 90 |
| Number of residents | 73 | 85.3 | 233.0 | 0.3 | 5.4 | 17.4 | 60.9 | 1,350.4 |
| Gross domestic product | 73 | 25,132 | 22,542 | 479 | 7144 | 14,501 | 42522 | 99,143 |

Note. SD=Standard deviation, MIN=Minimum, 25%=25$^{th}$ percentile, Med=Median, 75%=75$^{th}$ percentile, MAX=Maximum

Table 2. Number of institutions included in the statistical analyses for 17 different subject areas. The mean best paper rate/best journal rate is the mean best paper rate/best journal rate for the institutions within one subject area.

| Subject area | Number of institutions | Mean best paper rate | Mean best journal rate |
| --- | --- | --- | --- |
| Agricultural and Biological Science | 545 | 0.15 | 0.60 |
| Biochemistry, Genetics and Molecular Biology | 773 | 0.14 | 0.55 |
| Chemical Engineering | 167 | 0.14 | 0.54 |
| Chemistry | 511 | 0.13 | 0.62 |
| Computer Science | 381 | 0.14 | 0.34 |
| Earth and Planetary Sciences | 331 | 0.18 | 0.66 |
| Engineering | 644 | 0.14 | 0.41 |
| Environmental Science | 245 | 0.17 | 0.71 |
| Immunology and Microbiology | 217 | 0.16 | 0.65 |
| Materials Science | 412 | 0.13 | 0.55 |
| Mathematics | 379 | 0.14 | 0.45 |
| Medicine | 1231 | 0.17 | 0.58 |
| Neuroscience | 123 | 0.18 | 0.65 |
| Pharmacology, Toxicology and Pharmaceutics | 92 | 0.17 | 0.68 |
| Physics and Astronomy | 668 | 0.15 | 0.61 |
| Psychology | 70 | 0.20 | 0.61 |
| Social Sciences | 199 | 0.18 | 0.60 |

Table 3. Random-intercept models with a single cluster-covariate (subject area) as fixed effect for best paper rate

|  | Par | M$_0$ Est | CL | CU | M$_1$ International collaboration Est | CL | CU | M$_2$ Corruption perception index Est | CL | CU | M$_3$ Number of residents Est | CL | CU | M$_4$ Gross domestic product Est | CL | CU |
|---|---|---|---|---|---|---|---|---|---|---|---|---|---|---|---|---|
| *Fixed effects* | | | | | | | | | | | | | | | | |
| Intercept | $\beta_0$ | -2.03$^+$ | -2.07 | -2.00 | -1.57$^+$ | -1.62 | -1.54 | -2.07$^+$ | -2.17 | -2.03 | -2.07$^+$ | -2.10 | -2.04 | -2.03$^+$ | -2.06 | -2.00 |
| Covariate | $\beta_1$ | | | | 0.37$^+$ | 0.35 | 0.40 | 0.55$^+$ | 0.51 | 0.59 | -0.38$^+$ | -0.43 | -0.34 | 0.53 | 0.50 | 0.56 |
|  |  | | F | | | F | | | F | | | F | | | F | |
| Subject | $\beta_{2-18}$ | | 1,300.2$^*$ | | | 1,531.3$^*$ | | | 829.8$^*$ | | | 1,096.7$^*$ | | | 886.5$^*$ | |
| Subject × Cov | $\beta_{19-35}$ | | | | | 386.7$^*$ | | | 275.6$^*$ | | | 419.6$^*$ | | | 246.9$^*$ | |
| | | | | | | | | | | | | | | | | |
| *Random effects* | | | | | | | | | | | | | | | | |
| Intercept $u_{0j}$ | $\sigma^2_{u0|x}$ | 0.50$^\$$ | 0.47 | 0.53 | 0.34$^\$$ | 0.32 | 0.37 | 0.31$^\$$ | 0.29 | 0.34 | 0.43$^\$$ | 0.41 | 0.46 | 0.28$^\$$ | 0.26 | 0.30 |
| | | | | | | | | | | | | | | | | |
| ICC$_{Cov}$ | $\rho$ | 0.13 | | | 0.06 | | | 0.08 | | | 0.11 | | | 0.07 | | |
| R$^2_{u.Cov}$ | | 0.00 | | | 0.31 | | | 0.37 | | | 0.14 | | | 0.45 | | |
| | | | | | | | | | | | | | | | | |
| Deviance | | | 110,965 | | | 101,394 | | | 106,665 | | | 104,948 | | | 106,728 | |
| BIC | | | 112,001 | | | 101,658 | | | 106,930 | | | 105,213 | | | 106,992 | |

Note. Par = parameter, Est = estimated parameter, CL = lower limit of the 95% confidence interval, CU = upper limit of the 95% confidence interval, F = F-test, ICC = intra-class correlation conditioned on the covariate, R$^2_{u.cov}$ = coefficient of determination
$^*$ p< .05, F(16; 6,889)
$^+$ p< .05, t (6,872)
$^\$$ p< .05, Wald z-test

Table 4. Random-intercept models with a single covariate (subject area) as fixed effect for best journal rate

|  |  | $M_0$ | | | $M_1$ International collaboration | | | $M_2$ Corruption perception index | | | $M_3$ Number of residents | | | $M_4$ Gross domestic product 2011 | | |
| --- | --- | --- | --- | --- | --- | --- | --- | --- | --- | --- | --- | --- | --- | --- | --- | --- |
|  | Par | Est | CL | CU | Est | CL | CU | Est | CL | CU | Est | CL | CU | Est | CL | CU |
| *Fixed effects* | | | | | | | | | | | | | | | | |
| Intercept | $\beta_0$ | $-0.23^+$ | $-0.27$ | $-0.19$ | $0.43^+$ | $0.39$ | $0.47$ | $-0.24^+$ | $-0.27$ | $-0.20$ | $-0.24^+$ | $-0.27$ | $-0.20$ | $-0.22^+$ | $-0.25$ | $-0.18$ |
| Covariate | $\beta_1$ | | | | $0.53^+$ | $0.51$ | $0.55$ | $0.67^+$ | $0.63$ | $0.70$ | $-0.39^+$ | $-0.43$ | $-0.35$ | $0.68^+$ | $0.64$ | $0.71$ |
|  |  | F | | | F | | | F | | | F | | | F | | |
| Subject | $\beta_{2\text{-}18}$ | $14{,}424.6^*$ | | | $10{,}873.2^*$ | | | $13{,}757.6^*$ | | | $13{,}525.5^*$ | | | $13{,}517.9^*$ | | |
| Subject × Cov | $\beta_{19\text{-}35}$ | | | | $799.4^*$ | | | $1{,}256.2^*$ | | | $1{,}022.6*$ | | | $1{,}177.4^*$ | | |
| *Random effects* | | | | | | | | | | | | | | | | |
| Intercept $u_{0j}$ | $\sigma^2_{u0|x}$ | $0.78^\$$ | $0.73$ | $0.83$ | $0.53^\$$ | $0.50$ | $0.57$ | $0.46^\$$ | $0.44$ | $0.50$ | $0.63^\$$ | $0.59$ | $0.67$ | $0.41^\$$ | $0.39$ | $0.44$ |
| $ICC_{Cov}$ | $\rho$ | | | | $0.18$ | | | $0.13$ | | | $0.11$ | | | $0.15$ | | $0.10$ |
| $R^2_{u.Cov}$ | | | | | $0.00$ | | | $0.32$ | | | $0.41$ | | | $0.20$ | | $0.47$ |
| Deviance | | $233{,}806$ | | | $201{,}945$ | | | $212{,}762$ | | | $217{,}987$ | | | $213{,}690$ | | |
| BIC | | $233{,}942$ | | | $202{,}209$ | | | $213{,}027$ | | | $218{,}244$ | | | $213{,}955$ | | |

Note. Par = parameter, Est = estimated parameter, CL = lower limit of the 95% confidence interval, CU = upper limit of the 95% confidence interval, F = F-test, ICC = intra-class correlation conditioned on the covariate, $R^2_{u.cov}$ = coefficient of determination

$^*$ $p < .05$, $F(16; 6{,}889)$
$^+$ $p < .05$, $t(6{,}872)$
$^\$$ $p < .05$, Wald z-test

Table 5. Correlations (Pearson) among the country-level covariates (N = 73 countries)

| Variable | Corruption perception index | Number of residents | Gross domestic product |
| --- | --- | --- | --- |
| Corruption perception index | 1.00 | | |
| Number of academic institutions | .08 | 1.00 | |
| Gross domestic product | .81* | -.22 | 1.00 |

* $p < .05$

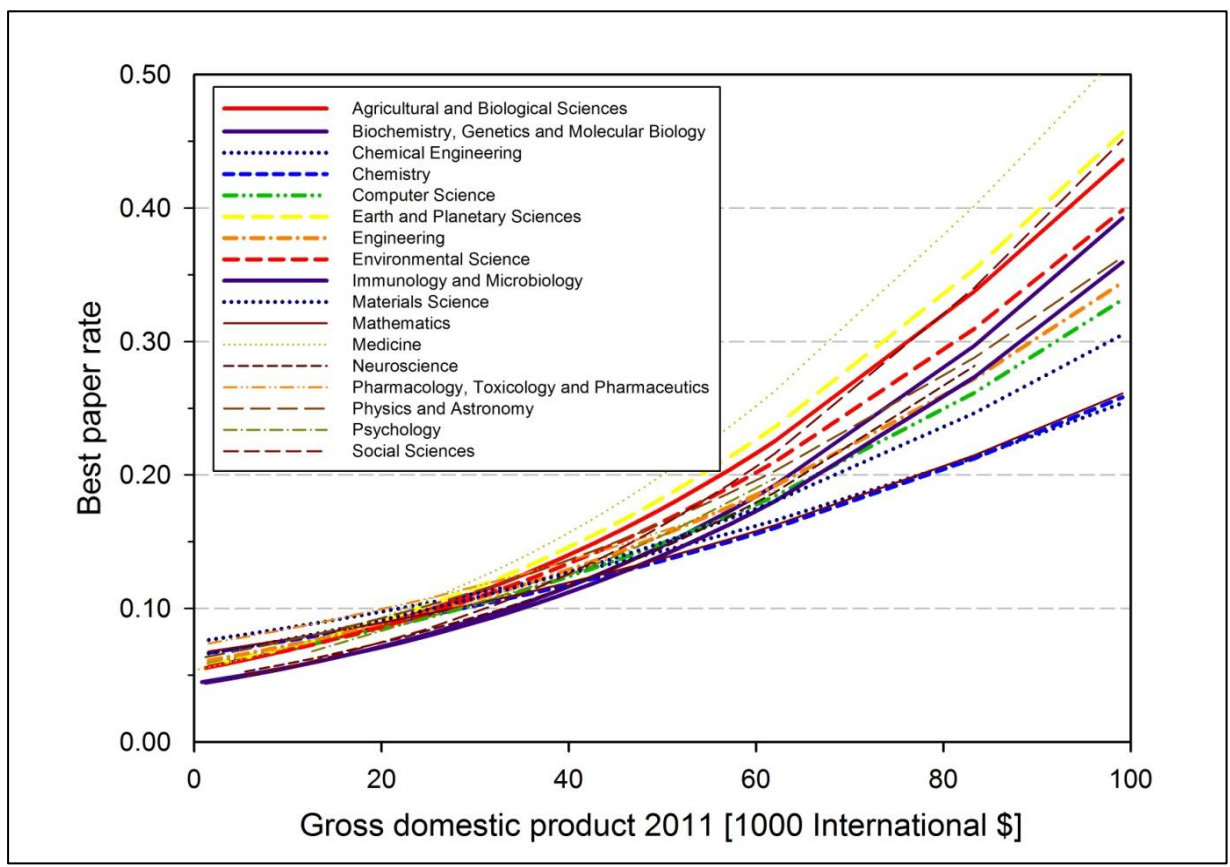

Figure 1. Predicted best paper rates in dependence on the gross domestic product by each subject area (without considering differences between institutions)